\documentclass[aps,twocolumn,superscriptaddress,pra,showpacs]{revtex4}
\usepackage{epsfig}
\usepackage{amssymb}
\usepackage{amsmath}
\usepackage{graphicx}   
\usepackage{bm}   
\usepackage{float,times}
\usepackage{epsfig}
\usepackage{theorem}
\usepackage{wrapfig}
\usepackage{picinpar}
\usepackage[normalem]{ulem} 
\usepackage{color}

\newcommand{\ket}[1]{|#1 \rangle}
\newcommand{\bra}[1]{\langle #1|}

\begin{document}

\title{Subspace Confinement: How Good is your Qubit?}
    \author{Simon J. Devitt}
    \affiliation{Centre for Quantum Computation, Department of Applied
      Mathematics and Theoretical Physics, University of Cambridge, Wilberforce
      Road, Cambridge CB3 0WA,
    United Kingdom}
    \affiliation{Centre for Quantum Computing Technology, Department of Physics,
    University of Melbourne, Victoria, Australia}
    \email{devitt@physics.unimelb.edu.au}
   \author{Sonia G. Schirmer}
    \affiliation{Centre for Quantum Computation, Department of Applied
      Mathematics and Theoretical Physics, University of Cambridge, Wilberforce
      Road, Cambridge CB3 0WA,
    United Kingdom}
    \author{Daniel K. L. Oi}
    \affiliation{Centre for Quantum Computation, Department of Applied
      Mathematics and Theoretical Physics, University of Cambridge, Wilberforce
      Road, Cambridge CB3 0WA,
    United Kingdom}
     \affiliation{SUPA, Department of Physics, University of Strathclyde,
      Glasgow G4 0NG, United Kingdom}
     \author{Jared H. Cole}
    \affiliation{Centre for Quantum Computing Technology, Department of Physics,
    University of Melbourne, Victoria, Australia}
    \author{Lloyd C.L. Hollenberg}
    \affiliation{Centre for Quantum Computing Technology, Department of Physics,
    University of Melbourne, Victoria, Australia}
    

\begin{abstract}
  The basic operating element of standard quantum computation is the qubit, an
  isolated two-level system that can be accurately controlled, initialized and
  measured.  However, the majority of proposed physical architectures for
  quantum computation are built from systems that contain much more
  complicated Hilbert space structures.  Hence, defining a qubit requires the
  identification of an appropriate controllable two-dimensional sub-system.
  This prompts the obvious question of how well a qubit, thus defined, is
  confined to this subspace, and whether we can experimentally quantify the
  potential leakage into to states outside the qubit subspace.  In
  this paper we demonstrate that subspace leakage can be quantitatively
  characterized using minimal theoretical assumptions by examining the Fourier
  spectrum of the oscillation experiment.
\end{abstract}
\maketitle

\section{Introduction}

The issue of subspace confinement for qubit systems is fundamental to the primary 
operating assumptions of quantum processors.  The concepts of universality, quantum 
gate operations, algorithms, error correction and fault-tolerant computation hinge 
on the precept that the fundamental quantum system is an isolated, controllable, 
two-dimensional system (qubit).    

It is well known that most of the physical realizations of qubits are
in fact multi-level quantum systems, which can theoretically be confined to a
two-dimensional (qubit) subspace. Important examples range from
super-conducting qubits~\cite{SC1,SC2,SC3} to atomic systems such as
cavity-coupled color centers~\cite{NV1,NV2,NV3} and ion
traps~\cite{ion1}. In the former systems, a qubit is generally
defined as the subspace (of the full Hilbert space) spanned by the two lowest
energy states in an arbitrarily shaped potential such as the washboard
potential of current-biased~Josephson Junctions~\cite{wash1,wash2}.  However, the
potential number of valid quantum states within each well is not limited to
two, and quantum gates, especially if sub-optimally implemented, may
inadvertently populate other confined states.  
Similarly in ion trap systems, a qubit is usually defined by two electronic states of an ion, either two
hyperfine levels or a ground state and a meta-stable excited state,
but once again there exist many other electronic states.  Hence a 
more stringent definition of a qubit would consist of a two-level quantum system 
with classical control confined to the unitary group $SU(2)$.

The ability to initialize, operate and measure completely within the two-level 
subspace representing  ``the qubit'' is vital to the successful operation of any 
large scale device constructed from such quantum systems.  Standard quantum error 
correction protocols (QEC)~\cite{QEC1,QEC2,QEC3} generally assume that the qubit 
system is precisely confined to the two-level subspace and that all quantum gates 
operate only on the qubit degrees of freedom.  If poor control or environmental 
influences inadvertently results in non-zero population of higher levels, 
leakage correction protocols are necessary.   

The issue of subspace leakage in quantum processing has been addressed in depth 
from the standpoint of error correction.  Work by Lidar~\cite{Lidar1,Lidar2} 
examined the construction of Leakage Reduction Units (LRU's), which use modified 
pulsing techniques to ensure that any unitary dynamics outside the qubit subspace 
can be compensated for which has been adapted specifically for super-conducting systems~\cite{mass}.  Another type of LRU's uses quantum teleportation%
~\cite{tele1} to map a multi-level quantum state back to a freshly 
initialized two-level qubit.  Finally, active detection such as non-demolition 
measurements (which detect population in non-qubit states without discriminating 
between the qubit states) can be performed on the system~\cite{det1,det2,det3,det4}.  
If an out-of-subspace detection event occurs the leaked qubit is re-initialized 
or replaced.  The inclusion of LRU's based on teleportation has been investigated 
within the context of fault-tolerant quantum computation~\cite{alifredis} and shows 
that, in principle, the inclusion of leakage protection does not adversely affect 
large scale concatenated error correction. 
   
Although these schemes are viable methods to detect and correct for improperly 
confined qubit dynamics, they can be cumbersome to implement and many systems
admit, in principle, sufficiently confined Hamiltonian dynamics so that leakage 
could be expected to be heavily suppressed.  For example, for ion-trap qubits 
controlled by lasers, leakage to other ionic states can be made negligible 
by employing very finely tuned lasers and sufficiently long (and 
possibly optimally tailored) control pulses.  Advances in qubit engineering may 
therefore allow us to eliminate or at least substantially reduce the need for 
laborious leakage detection/prevention schemes in many cases, provided that we 
can experimentally ascertain sufficiently high confinement of manufactured qubits
under classically controlled Hamiltonian dynamics.

In this paper we present a simple generic protocol to estimate qubit confinement,
or more precisely, establish bounds on the subspace leakage rates, for ``quality 
control'' purposes.  The main goal is to allow us to empirically detect inferior 
qubits by using readily obtainable experimental data to derive tight bounds on the 
subspace leakage of the system.  This protocol would represent one of the first 
steps towards full system characterization~\cite{char1,char2,char3,char4}.

Section~\ref{sec:prelim} briefly outlines the basic assumptions with respect to 
the measurement and control model and the motivation for the proposed protocol.
Section~\ref{sec:leakage} discusses the basic mathematical properties of 
qubit oscillation data and shows how a minimal amount of information obtained 
from the oscillation 
spectrum can be used to derive empirical bounds on the subspace leakage 
rate, and that these bounds are very tight for the high quality qubits required 
for practical quantum computation.  In section~\ref{sec:discrete}, the effects 
of finite sampling are considered and studied using numerical simulations.  
Section~\ref{sec:efficiency} compares the efficiency of bounding confinement 
using the proposed scheme versus alternative approaches such as detection of 
imperfect confinement by identifying additional transition peaks within the 
Rabi spectrum.  Finally, section~\ref{sec:decohere} briefly examines the effects 
of decoherence.

\section{Motivation and Preliminaries}
\label{sec:prelim}

Estimation of qubit confinement represents one of the first major steps in full 
qubit characterization.  Therefore, the protocol should not be predicated on the 
availability of sophisticated measurements or control, and should be amenable to
automation so that it could be used in conjunction with a potentially automated 
qubit manufacturing process. The bounds on the subspace leakage will be based on 
the observable qubit evolution under an externally controlled driving Hamiltonian.  
We assume that our classical control switches on the single qubit 
dynamics and that the governing Hamiltonian is piecewise constant in time.  Hence the 
Hamiltonian
induces the unitary operator $U=e^{-iHt}$, with $\hbar = 1$.  

Although this assumption
may not be applicable to all systems, e.g., systems subject to ultra-fast
tailored control pulses, it is not as restrictive as it might appear.  It
is generally be valid for systems such as quantum dots or Josephson
junctions subject to external potentials created by voltage gates if the
gate voltages are (approximately) piecewise constant.  It is also a good 
approximation for systems subject to time-dependent fields such as
laser pulses in a regime where the rotating wave approximation (RWA) is
valid and the pulse envelopes can be approximated by square-waves.  In this 
case, the Hamiltonian relevant for our purposes is the (piecewise
constant) RWA Hamiltonian determined by the amplitudes, detunings and
possibly phases of the control pulses.  This model can even 
be valid for other pulse shapes if the Hamiltonian is taken to be an 
average Hamiltonian describing the effective dynamics on a certain time scale 
(beyond which we do not resolve the time-dependent dynamics).  However, 
the main focus of the paper is not when the dynamics of a system can be 
modeled in this way, but rather how to assess subspace confinement for 
systems where this model of the dynamics is valid.

Assuming the effective control-dependent Hamiltonian $H = H[\vec{f}]$ is constant for $0\leq t\leq t_k$,
where $\vec{f}$ is the classical ``control knob" parameter, 
the evolution during this time period is given by the unitary operator $U(t) = e^{-iHt}$.  
Although $H$ will generally depend on control inputs, 
we shall omit this dependence in the following for notational convenience.  The driven 
system generally undergoes coherent oscillations, which are often
referred to as Rabi oscillations, especially for optically driven systems
in the RWA regime.  Although our model is not limited to these systems,
we shall use the terms coherent oscillations and Rabi oscillations 
interchangeably throughout this paper.

The measurement model assumed is crucial to the relevance of the protocol.
Some standard measurement models in quantum computation assume the ability to
detect both the $|0\rangle$ and $|1\rangle$ states independently (such as SET
detectors in solid state designs~\cite{SS1,SS2,SS3}). In this case,
estimating subspace leakage is fairly straightforward and requires only
repeated measurement of the system while undergoing evolution.  The leakage is
simply given by the deviation of the cumulative probability of measuring
$|1\rangle$ or $|0\rangle$ from unity.  However, this measurement model is not
realistic for the majority of proposed systems. 

Color centers and
ionic qubits use externally pumped transitions to discriminate between a light
state ($\equiv |0\rangle$) and other ``dark'' states, while readout in
super-conducting systems~\cite{SCmeas1,SCmeas2} involves lowering a potential
barrier such that only one of the qubit states can leak to an external
detection circuit.  The measurement outcome of the indirectly probed state is
inferred from the non-detection of the directly measured state and for such
measurement models estimating confinement is more complicated.  Hence 
this paper utilizes the latter model in order to quantify confinement.  It should be noted 
that we are not considering the concept of weak measurement, in each case we assume that 
the measurement of the system causes a full POVM collapse of the wavefunction.
We also assume that the measurement apparatus has been sufficiently
characterized.  In order to to successfully implement computation, readout
fidelity should ideally be of the same order as general systematic and decoherence errors.  
Therefore, characterization is initially required to ascertain the error rate associated with 
measurement which can then be incorporated into calculations of confinement.

\begin{widetext}

\begin{figure}[ht]
\includegraphics[width=0.9\textwidth]{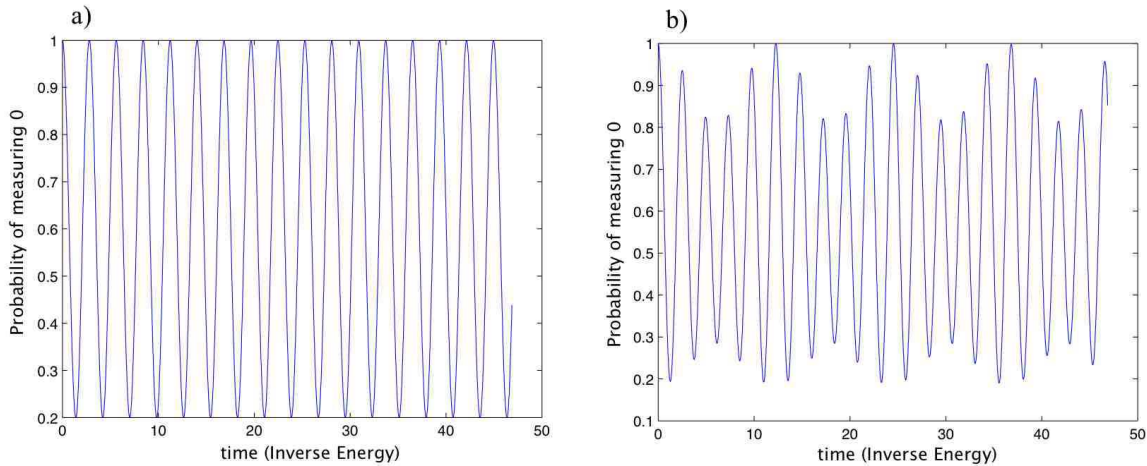}
\caption{Modulations in the Rabi oscillations of a three-level system driven by 
the Hamiltonians, $H_m$ and $H_n$.
Fig.~b) provides 
clear evidence that this system is not a qubit, while Fig.~a) appears to show perfect confinement.  
However the analysis in the following sections will show that the subspace 
confinement for the system in Fig. a)
is also not sufficient 
for large-scale QIP applications.}
\label{fig:h1}
\end{figure}

\end{widetext}

Strong non-qubit transitions can still be identified directly via modulations in 
the Rabi oscillations data as shown in Fig.~\ref{fig:h1}b for a three-state system
evolving under the trial Hamiltonian
\begin{equation}
 H_m = \begin{pmatrix} 
	0 & 1 & 0.5 \\ 
	1 & 1 & 0 \\
	0.5 & 0 & 1.5
	\end{pmatrix}.
	\label{eq:ham1}
\end{equation}
However, the Rabi oscillation data for the modified three-state Hamiltonian,
\begin{equation}
 H_n = \begin{pmatrix} 
	0 & 1 & 0.01 \\ 
	1 & 1 & 0 \\
	0.01 & 0 & 1.5
	\end{pmatrix},
	\label{eq:ham2}
\end{equation} 
depicted in Fig.~\ref{fig:h1}a shows that an apparent lack of modulations in the 
Rabi oscillation data is not proof of perfect confinement, and that quantitative 
measures of confinement or subspace leakage and experimental protocols are needed.

\section{Estimation of subspace leakage}
\label{sec:leakage}

By defining the projection operator onto a two dimensional subspace,
$\Pi=\ket{0}\bra{0}+\ket{1}\bra{1}$, subspace leakage is given by,
\begin{equation}
\epsilon = 1-\text{Tr}[\Pi\rho],
\end{equation} 
with $\rho = U^{\dagger}(t)\ket{0}\bra{0}U(t)$.
Unfortunately, we cannot calculate $\epsilon$ directly without knowledge of
the Hamiltonian.  However, we can estimate subspace leakage experimentally
from standard Rabi oscillation data.
\subsection{Perfect confinement}
\label{sec:perfect}

Consider a general $N$-level system undergoing coherent evolution via a driving 
Hamiltonian $H_N$ in the closed system case of no environmental decoherence.  
If confinement under this Hamiltonian is perfect, $H_N$ has 
a direct sum decomposition,
\begin{equation}
H_N  = H_{2 \times 2} \oplus H_{(N-2) \times (N-2)}
\label{eq:direct}
\end{equation}
where $H_{2 \times 2}$ represents the control Hamiltonian confined to the qubit 
subspace, $\mbox{span}\{\ket{0},\ket{1}\}$, the state $\ket{0}$ being defined 
by the measurement, and the excited state $\ket{1}$ by the allowed transition.
For our measurement model the observed Rabi oscillations have the functional 
form $f(t) = |\bra{0} U_N(t) \ket{0}|^2$.  As there is no coupling between the state 
$\ket{0}$ and states outside the $H_{2\times 2}$ subspace, we can expand $f(t)$ 
by diagonalizing $U_{2\times 2}(t) = \exp(-iH_{2\times 2}t)$
\begin{equation}
\begin{aligned}
f(t) &= |\bra{0} A^{\dagger}\text{diag}\{e^{-i\lambda_0 t},e^{-i\lambda_1 t}\}A\ket{0}|^2\\
     &= ||c_0|^2e^{-i\lambda_0 t}+|c_1|^2e^{-i\lambda_2 t}|^2 \\
     &= |c_0|^4+|c_1|^4 + |c_0|^2|c_1|^2 (e^{i\omega_{01} t} + e^{-i\omega_{01}t})
\end{aligned}
\end{equation}
where $U_{2\times 2}(t) = A^{\dagger}e^{-iH_dt}A$, 
$A\ket{0}=c_0\ket{0}+c_1\ket{1}$, $\omega_{01}=\lambda_0-\lambda_1$ and 
$\{\lambda_j\}$ are the eigenvalues of $H_{2\times 2}$.  For perfect confinement, $H_{2\times 2}$ 
induces coherent oscillations between the two qubit levels at a Rabi frequency 
given by the difference in the eigenvalues.  Taking the Fourier transform of 
$f(t)$ gives
\begin{equation}
\begin{aligned}
F(\omega) = \mbox{FT}[f(t)] &= (|c_0|^4+|c_1|^4)\delta(\omega) \\
+ |c_0|^2|c_1|^2\delta(\omega-&\omega_{01}) 
+ |c_0|^2|c_1|^2\delta(\omega+\omega_{01}).\\
\end{aligned}
\end{equation}
Conservation of probability (total population) thus implies $(|c_0|^2+|c_1|^2)^2 
=|c_0|^4+|c_1|^4+2|c_0|^2|c_1|^2=1$, and hence the heights of the two Fourier peaks
for perfect confinement will satisfy the relation $h_0+2h_{0,1} = 1$, where $h_0 = 
|c_0|^4+|c_1|^4$ and $h_{0,1} = |c_0|^2|c_1|^2$.   

\subsection{Imperfect confinement}
\label{sec:imper}

If the system experiences leakage to states outside the qubit subspace then the
corresponding control Hamiltonian $H_N$ can no longer be reduced to a direct sum 
representation~(\ref{eq:direct}) but it can be diagonalized $H_d = \text{diag} 
[\{\lambda_j\}]$, $\{\lambda_j\}$ being the eigenvalues of $H_N$, and the propagator 
$U_N(t)$ expressed as $U_N(t) = A^{\dagger}e^{-iH_dt}A$.  The Rabi data is now a linear 
superposition of multiple oscillations corresponding to different transitions of 
the $N$-level system 
\begin{equation}
\begin{aligned}
 f(t) & = |\bra{0}A^{\dagger}e^{-iH_dt}A\ket{0}|^2 \\
      & = \bigg{|} \sum_{a=0}^{N-1} |c_a|^2 e^{-i \lambda_at}\bigg{|}^2 
        = \sum_{a,b}|c_a|^2|c_b|^2e^{-i(\lambda_a-\lambda_b)t}
\end{aligned}
\end{equation}
and the corresponding peak heights in the Fourier spectrum can be expressed in 
terms of the expansion co-efficients, $A\ket{0} =  \sum_{a=0}^{N-1}c_a\ket{a}$, as,
\begin{equation}
\begin{aligned}
 h_0    &= \sum_{a=0}^{N-1}|c_a|^4, \quad \quad
 h_{a,b} = |c_a|^2|c_b|^2, \quad \quad a \neq b.
\end{aligned}
\label{eq:h0}
\end{equation}
Conservation of probability leads to
\begin{equation}
\begin{aligned}
1 = \bigg{(}\sum_{a=0}^{N-1}|c_a
& |^2\bigg{)}^2 =  \sum_{a=0}^{N-1} |c_a|^4 + \sum_{a \neq b}|c_a|^2|c_b|^2 \\
& = h_0 + \sum_{a \neq b}h_{a,b}.
\end{aligned}
\label{eq:identity}
\end{equation}
Imperfect confinement implies $h_0+2h_{0,1}<1$.  We see from this analysis 
that the subspace leakage $\epsilon$ is determined by the cumulative amplitudes of 
all non-qubit states for a given eigenstate of $H_N$, which can be calculated from 
{\em all} the peak heights in the Fourier spectrum,
\begin{equation}
\epsilon = \sum_a \sqrt{\frac{h_{a,b}h_{a,c}}{h_{b,c}}}, \quad b,c \neq a.
\end{equation}
However, exact calculation of 
$\epsilon$ requires identification of all peaks in the Fourier spectrum and knowledge of 
which peak corresponds to each trasition.  It is therefore desirable to 
derive bounds on the subspace leakage that only involve a few dominant and
thus easily identifiable Fourier peaks.

\subsection{Bounds on subspace leakage}
\label{sec:bounds}

We can derive upper and lower bounds on $\epsilon$ using only the heights of the 
primary spectral peaks $h_0$ and $h_{0,1}$.
\begin{equation} \label{eq:eps1}
\begin{aligned} 
h_0+2h_{0,1} &= (|c_0|^2+|c_1|^2)^2 + \sum_{a\neq 0,1}|c_a|^4 \\
            &= \text{Tr}[\Pi \rho]^2 + \sum_{a\neq 0,1}|c_a|^4 \\
            &= (1-\epsilon)^2 + \sum_{a\neq 0, 1}|c_a|^4.
\end{aligned}
\end{equation}
Provided $\sum_{a\neq 0,1}|c_a|^4 \ll 1$, i.e., subspace leakage is reasonably small, 
we obtain a tight lower bound for $\epsilon$ as a function of only the two major 
peak heights:
\begin{equation}
\begin{aligned}
h_0+2h_{0,1} &\geq (1-\epsilon)^2 \\
\therefore \quad \epsilon &\geq 1-\sqrt{h_0+2h_{0,1}}.
\end{aligned}
\end{equation}

The upper bound for $\epsilon$ can also be calculated quite easily.  Recall that
\begin{equation} \label{eq:eps2}
\begin{aligned}
 \epsilon^2 & = \bigg{(}\sum_{a\neq 0,1} |c_a|^2 \bigg{)}^2 
            = \sum_{a \neq 0,1}|c_a|^4 + \sum_{a,b>1, a\neq b}|c_a|^2|c_b|^2 \\
            & \ge \sum_{a \neq 0,1}|c_a|^4. 
\end{aligned}
\end{equation}
Comparison with (\ref{eq:eps1}) thus immediately yields
\begin{equation} \label{eq:eps} 
  h_0+2h_{0,1} \leq (1-\epsilon)^2+\epsilon^2 = 1-2\epsilon + 2\epsilon^2,
\end{equation}
which can be solved for $\epsilon$
\begin{equation} \label{eq:upper}
  \epsilon \leq \frac{1}{2}(1-\sqrt{2h_0+4h_{0,1}-1}). 
\end{equation}
The other solution to Eq.~(\ref{eq:eps}) is invalid as a bound due to the 
asymptotic behavior of both the upper and lower bound
\begin{equation}
\begin{aligned}
&\lim_{(h_0+2h_{0,1}) \rightarrow 1} \text{min}(\epsilon) = 0, \\
&\lim_{(h_0+2h_{0,1}) \rightarrow 1} \text{max}(\epsilon) = 0.
\end{aligned}
\end{equation}
Since the second term in (\ref{eq:eps2}) represents the heights of all the Fourier 
peaks \emph{not} associated with the $\ket{0} \leftrightarrow \ket{1}$, $\ket{0} 
\leftrightarrow\ket{a}$ or $\ket{1} \leftrightarrow \ket{a}$ transitions, 
for $\ket{a} \neq \ket{1}$.  For a well confined system this is a very 
small correction to $\epsilon^2$, consequently the bound is again strong.

Therefore, the subspace leakage $\epsilon$ is bounded above and below by
\begin{equation} \label{eq:bounds}
 1-\sqrt{h_0+2h_{0,1}} \leq \epsilon \leq \frac{1}{2}(1-\sqrt{2h_0+4h_{0,1}-1}).
\end{equation}
Note that this double inequality involves only the two main peaks in the Fourier
spectrum, i.e., we can bound the subspace leakage \emph{without} determining the heights of all 
peaks.  

For the trial Hamiltonians~(\ref{eq:ham1}) and (\ref{eq:ham2}) we obtain the
following bounds
\begin{equation}
\begin{aligned}
0.0497 \leq & \epsilon_{H_m} \leq 0.0511, \\
3.9754 \times 10^{-4} \leq & \epsilon_{H_n} \leq 3.9762\times 10^{-4},
\end{aligned}
\end{equation}
while the actual values of $\epsilon_{H_m}$ and $\epsilon_{H_n}$ are
\begin{equation}
\epsilon_{H_m} = 5.11\times 10^{-2}, \quad \quad \epsilon_{H_n} = 3.9762\times 10^{-4}.
\end{equation}
In both cases the upper bound for $\epsilon$ equals the actual value of $\epsilon$.  
This is due to the fact that both systems are of dimension three, and when estimating 
$\text{max}(\epsilon)$ we neglected terms of the form
\begin{equation}
 \sum_{(a,b) \neq (0,1),  a\neq b}|c_a|^2|c_b|^2,
\end{equation}
which naturally vanish for a three-level system.  

\begin{figure}[ht]
\includegraphics[width=0.5\textwidth]{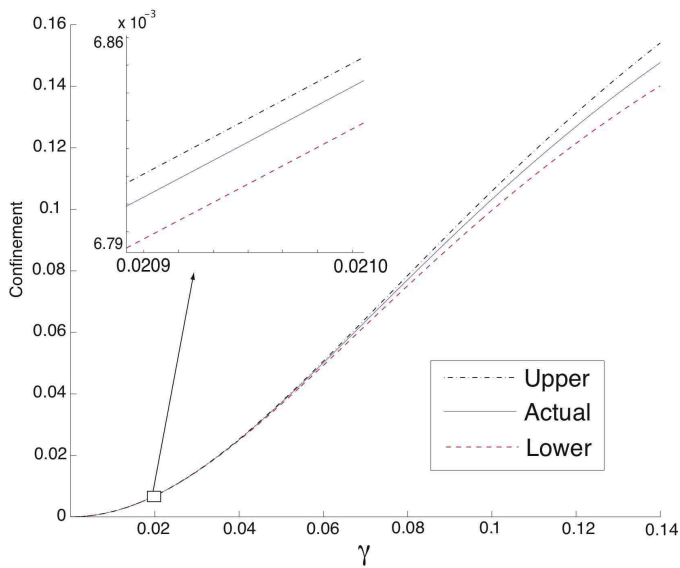}
\caption{Upper and lower bounds on $\epsilon$ for the four-level trial system 
governed by the Hamiltonian~(\ref{eq:H4}), characterized by a static coupling 
between the qubit states and a variable coupling $\gamma$ to two higher levels.  
As $\gamma \rightarrow 0$ the subspace leakage approaches 0 and the bounds for 
$\epsilon$ become more accurate.}
\label{fig:test}
\end{figure}

Fig.~\ref{fig:test} shows how the bounds~(\ref{eq:bounds}) for $\epsilon$ converge 
as confinement increases ($\gamma \rightarrow 0$) for the test Hamiltonian,
\begin{equation}
H_4 = \begin{pmatrix} 0 & 1 & \gamma & \gamma \\
                      1 & 1 & 0 & 0 \\
		      \gamma & 0 & 1.5 &0 \\
                      \gamma & 0 & 0 & 1.7
\end{pmatrix}.
\label{eq:H4}
\end{equation}

\section{Finite Sampling Fourier analysis}
\label{sec:discrete}

The previous section details how quantitative bounds on the subspace leakage can
be obtained, in principle, from the Fourier spectrum of the Rabi data.  However, 
to translate this method into a viable experimental protocol we need to consider
the effects of finite sampling and taking the discrete Fourier transform (DFT), 
which raises several issues.

First the Nyquist criterion for sampling~\cite{DFT} must be satisfied, i.e., to
avoid aliasing, some rough estimate of the Rabi period $T_{\text{Rabi}}$ is needed 
to guarantee that at least two sample points are chosen per oscillation period, 
i.e., $\Delta t \leq T_{\text{Rabi}}/2$.  
The second issue that must be considered is the resolution of the Fourier spectrum.  
The frequency resolution $\Delta \omega$ is given by $\Delta \omega = 2\pi/t_{ob}$, 
with $t_{ob}$ the total observation time of the Rabi signal. 
If the control Hamiltonian induces a non-qubit transition with a frequency within 
$\Delta \omega$ of the primary peak then the DFT will combine the amplitudes for 
qubit and non-qubit transitions in the same frequency channel thus leading to an
overestimate of $h_{0,1}$ and hence qubit confinement.  To avoid such problems it 
is necessary to ensure that the total observation time $t_{ob}$ is long enough. 
Thus, some estimates of the system parameters are required, although these do not
need to be very accurate and will generally be known on theoretical or experimental
grounds.

Finally, the DFT has the property that a pure sinusoidal signal will approach a 
delta function if there is zero phase difference between the start and 
the end of the observed signal.  If this phase matching condition is not met then 
all frequency peaks will broaden.  Phase matching for system identification has 
already been addressed for the identification of single qubit control Hamiltonians 
in~\cite{char4} and we will follow the same approach, which essentially involves 
truncating the Rabi oscillation data at progressively greater values of $t_{ob}$ 
such as to maximize the trial function
\begin{equation}
 P(t_{ob}) = \frac{2 F(\omega_p)-F(\omega_p-1)-F(\omega_p+1)}{F(\omega_p-1)+F(\omega_p+1)},
\end{equation}
where $F(\omega)$ represents the amplitude of the Fourier Spectrum at frequency 
$\omega$ and $\omega_p$ represents the frequency of the maximum Fourier peak.  
The value of $t_{ob}$ where $P(t_{ob})$ is maximized represents the cut off time 
to the Rabi signal that produced the best phase matching for the DFT.  


To simulate real experiments we numerically propagate the initial state $\ket{0}$, under the 
Hamiltonian $H$,  
by $U(t_k)=\exp(-i t_k H)$ for discrete times $t_k=k \Delta t$ where $k=0,1,\ldots,K$
and $K\Delta t=t_{ob}$.  A single measurement at time $t_k$ is simulated by mapping 
the target state $U(t_k)\ket{0}$ to $\{\ket{0},\ket{1}\}$, where the probability of obtaining 
$0$ is given by $p_0=|\bra{0}U(t_k)\ket{0}|^2$; the ensemble average at a single time 
$t_k$ is determined by dividing the number of zero results by the total number of 
repeat experiments $N_e$.  For the following numerical simulations we shall use the trial 
Hamiltonians
\begin{equation} \label{eq:ham31}
  H_a = \begin{pmatrix} 
	0 & 1 & 0 & 0 & 0 \\ 
	1 & 1 & 0 & 0 & 0\\
	0 & 0 & 1.5 & 0 & 0 \\
	0 & 0 & 0 & 1.7 & 0 \\
	0 & 0 & 0 & 0 & 2
	\end{pmatrix},	
\end{equation}
and
\begin{equation} \label{eq:ham3}
  H_b = \begin{pmatrix} 
	0 & 1 & 0.01 & 0.005 & 0 \\ 
	1 & 1 & 0 & 0 & 0\\
	0.01 & 0 & 1.5 & 0 & 0 \\
	0.005 & 0 & 0 & 1.7 & 0 \\
	0 & 0 & 0 & 0 & 2
	\end{pmatrix}, 
\end{equation}
where $H_a$ represents a five-level system with a perfectly decoupled two-level 
subspace consisting of the two lowest energy states, while $H_b$ represents a 
five-level system with weak coupling between the qubit sub-manifold and two of 
the upper levels.  We only consider Hamiltonians that have couplings between 
the $\ket{0}$ state and higher levels, as this state is fixed by the measurement basis.  
We are therefore free to diagonalize the lower block of the Hamiltonian, which also 
helps to simplify the comparison between different systems.

The out-of-subspace coupling in $H_b$ was chosen such that 
the leakage from the qubit subspace $\epsilon \approx 7\times 10^{-4}$ is small 
(too small to cause noticeable modulations in the Rabi oscillations) yet 
significant (in fact above certain critical thresholds) for quantum computing 
applications.  The part of the Hamiltonian governing the qubit dynamics was 
chosen arbitrarily and is common to all the Hamiltonians examined within this paper
to maintain consistency between different simulations.  The 
accuracy of the protocol is not affected by the choice of single qubit dynamics. 

\subsection{Estimating uncertainty in leakage bounds}

Estimating uncertainties in the bounds for $\epsilon$ is crucial since for the 
majority of qubit systems it will be practically impossible to prove that the
evolution of the system under a given Hamiltonian is completely confined to the 
$SU(2)$ subspace, i.e., $\epsilon=0$.  Instead, in practice it is sufficient 
for quality control purposes to experimentally confirm that the leakage from 
the qubit subspace is below a threshold value where it can effectively be 
ignored, i.e., it is the upper bound $\text{max}(\epsilon)$ that is relevant.  
The accuracy of our estimate for $\text{max}(\epsilon)$ will be primarily 
limited by our ability to accurately determine the main peak heights $h_0$ and 
$h_{0,1}$ due to projection noise induced by the DFT.  

Quantifying this uncertainty is relatively straightforward.  Defining the noise
function $\nu(\omega)$ of the Fourier spectrum to be the amplitude $\nu(\omega)$ 
of each Fourier channel excluding $h_0 = F(0)$ and $h_{0,1} = F(\omega_p)$, the 
uncertainty in $h_0$ and $h_{0,1}$ is given by the standard deviation of the noise 
function $\delta h = \text{sd}[\nu(\omega)]$.  From this we can derive the
uncertainty associated with $\text{max}(\epsilon) \equiv \epsilon_u$.
\begin{equation}
\begin{aligned}
(\delta \epsilon_u)^2  
  &= \bigg{(}\frac{\partial \epsilon_u}{\partial h_0} \bigg{)}^2 (\delta h_0)^2
   + \bigg{(}\frac{\partial \epsilon_u}{\partial h_{01}}\bigg{)}^2 (\delta h_{0,1})^2 \\
   &+ 2\bigg{(}\frac{\partial \epsilon_u}{\partial h_0} \bigg{)}
   \bigg{(}\frac{\partial \epsilon_u}{\partial h_{01}}\bigg{)}\delta h_0 \delta h_{0,1}
  = \frac{3\delta h}{2\sqrt{2h_0+4h_1-1}}.
  \end{aligned}
\end{equation}

\begin{figure}[ht]
\includegraphics[width=0.45\textwidth]{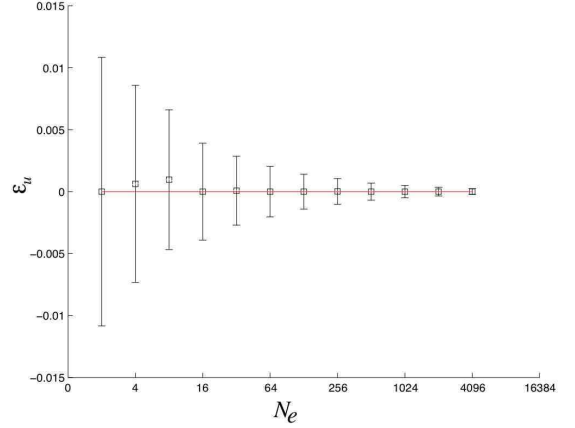}
\caption{Convergence of $\epsilon_u$ as the number of ensemble measurements, $N_e$,
is increased for a system governed by the Hamiltonian~(\ref{eq:ham3}), characterized
by perfect subspace confinement.  The solid line represents the actual value of 
$\epsilon_u(H_a)=0$. Note error bars should only extend to zero as $\epsilon_u(H_a) \geq 0$.}
\label{fig:q1}
\end{figure}

\begin{figure}[ht]
\includegraphics[width=0.45\textwidth]{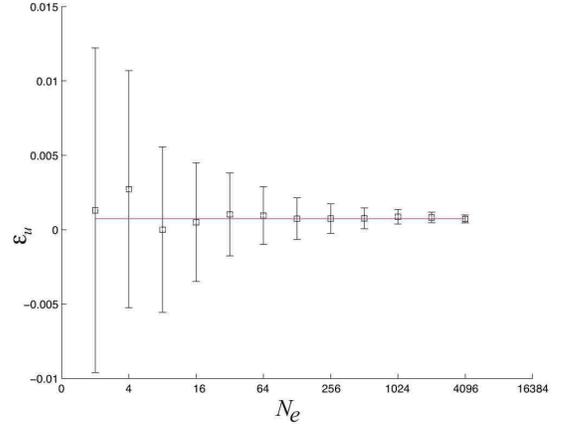}
\caption{Convergence of $\epsilon_u$ as the number of ensemble measurements, $N_e$,
is increased for the imperfectly confined system governed by the Hamiltonian~%
(\ref{eq:ham31}).  The solid line represents the actual value of $\epsilon_u(H_b) 
\approx 7 \times 10^{-4}$.  Note error bars should only extend to zero as 
$\epsilon_u(H_b) \geq 0$.}
\label{fig:q2}
\end{figure}

$\delta \epsilon_u$ can be reduced by increasing the number of ensemble measurements 
$N_e$ taken at each point in the Rabi cycle.  Figures~\ref{fig:q1} and \ref{fig:q2} 
show how the estimate for $\epsilon_u$ converges as $N_e$ is increased for the 
Hamiltonians~(\ref{eq:ham31}) and (\ref{eq:ham3}), respectively.  It should be noted 
that $\epsilon_u \geq 0$, hence for each plot the lower error bars should only 
extend to the zero point, but keeping the error bars symmetrical around the data 
point makes the convergence behavior clearer.  
For large values of $N_e$, $\epsilon_u$ converges to zero for the perfectly confined 
system governed by $H_a$ but the non-zero value $\approx 7\times 10^{-4}$ for the 
imperfectly confined system described by $H_b$.  The respective observation times 
for each Hamiltonian were chosen to be $t_{ob}= 30T_{\text{Rabi}}$ to ensure that all peaks
are resolved, i.e., there are no contributions from additional transitions present 
within $\Delta \omega$ of the primary peak.  

\subsection{Numerical tests of error bound accuracy}

\begin{figure}[ht]
\includegraphics[width=0.45\textwidth]{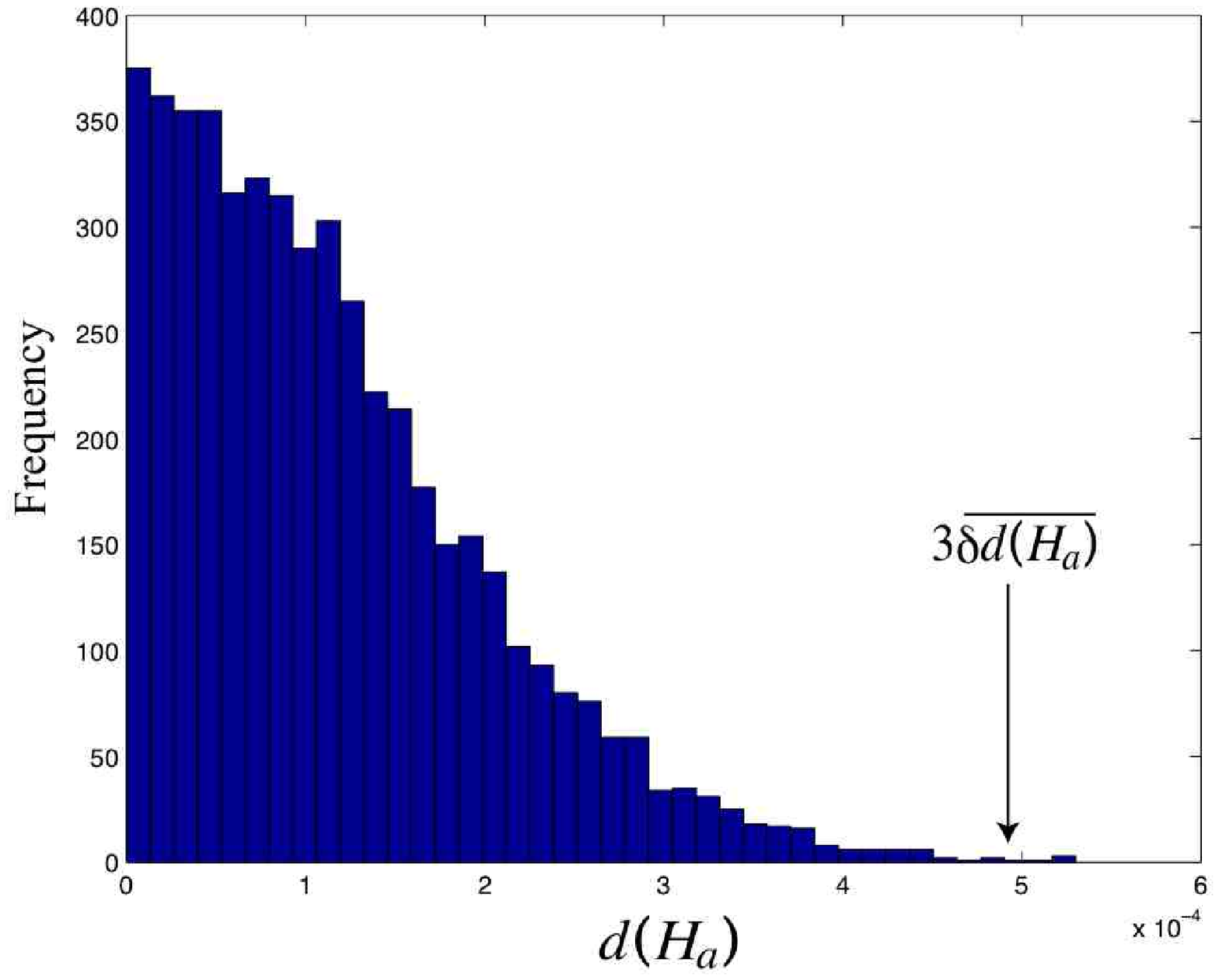}
\caption{Distribution of $d(H_a)$ for 5000 separate simulations. The average
  of the error, $3\overline{\delta  d(H_a)}$ is also shown, with approximately
  $99.9 \%$ found within $3\sigma$ of $d = 0$.}
\label{fig:q3}
\end{figure}

\begin{figure}[ht]
\includegraphics[width=0.45\textwidth]{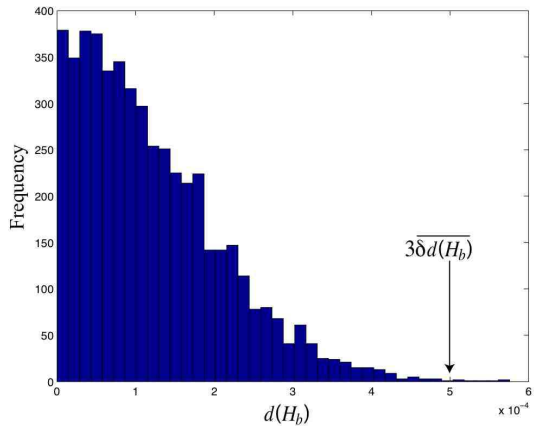}
\caption{Distribution of $d(H_b)$ for 5000 separate simulations. The average
  of the error, $3\overline{\delta d(H_b)}$ is also shown, with approximately $99.8
  \%$ found within $3\sigma$ of $d = 0$.}
\label{fig:q4}
\end{figure}

To test the overall accuracy of the uncertainty estimates for $\epsilon_u$ we can
expect to obtain from realistic Rabi oscillation data, we calculated the 
distance between the simulated value, $\epsilon_u$, and the analytical value, $\epsilon_u'$, calculated 
directly from the Hamiltonian using Eq. \ref{eq:upper} as,
\begin{equation}
  d(H_k) = |\epsilon_u(H_k)-\epsilon_u'(H_k)|,
\end{equation}
where $k \in [a,b]$ and $\delta d(H_k)$ is the error in $d$ resulting from the error associated with 
estimating $\epsilon_u(H_k)$.
We first calculated the distance $d(H_k)$ and $\delta d(H_k)$ for 5000 simulated 
runs of two known trial Hamiltonians ($H_a$ and $H_b$) with $\epsilon_u'(H_a)=0$ 
and $\epsilon_u'(H_b) \approx 7\times 10^{-4}$, respectively.  The distributions 
of $d(H_k)$ for $H_a$ and $H_b$ (with $N_e = 1024$ and $t_{\rm ob}$ as in Figs~%
\ref{fig:q1} and \ref{fig:q2} are shown in Figs~\ref{fig:q3} and \ref{fig:q4},
respectively.  The average error $3\overline{\delta d(H_i)}$, $i \in[a,b]$, was given by 
$3\overline{\delta d(H_a)} \approx 4.92\times 10^{-4}$, encompassing $99.9 \%$ of the 
data, and $3\overline{\delta d(H_b)} \approx 5.02 \times 10^{-4}$, encompassing $99.8 
\%$ of the data, respectively.  

Next we examined how the protocol behaves when simulating a large number of 
randomly selected multi-level Hamiltonians.  For these simulations we choose 
$N$-level Hamiltonians of the form
\begin{equation}
 H_N = \sum_{k=0}^{9} E_k|k\rangle\langle k| + |0\rangle\langle1| 
       + \sum_{k=2}^9 a_k|0\rangle\langle k|+\text{h.c.}
\end{equation}
with $\{E_k\} \equiv \{0,1,1.5,2,2.4,2.5,2.9,3,3.3,4\}$.  The vector $\vec{a} = [a_2,
\ldots, a_9]$ was then chosen at random in two stages.  First the dimensionality 
of $\vec{a}$ is randomly selected, allowing the Hamiltonian to coherently drive
any multi-level system, $N \in [2,3,..,10]$.  The non-zero coupling values 
were then randomly assigned such that each element of $\vec{a}$ was approximately two
orders of magnitude less than the qubit coupling term to ensure that all of the
multi-level systems had high confinement.

We randomly generated 5000 of these Hamiltonians and $d(H_k)=|\epsilon_u(H_k)-\epsilon_u'(H_k)|$ 
was calculated.  The average (analytical) value of $\epsilon_u'(H_k)$ for these 5000 trial
Hamiltonians was found to be $\overline{\epsilon_u'(H_k)} = 1.68\times 10^{-4}$.  We then 
examined the ratio,
\begin{equation}
 R = \frac{\text{Num} \{ (d(H_k)-3\delta d(H_k) \leq 0) \}}{5000},
\end{equation}
indicating the percentage of successful estimates of the subspace leakage within 
$3\sigma$.  This ratio was calculated to be $R = 99.9 \%$, with the confinement 
estimates being outside the error bounds for only three of the randomly generated 
Hamiltonians.  

These results are consistent with the expectation that approximately $99.7 \%$ 
of the data should lie within $3\sigma$ of the mean and demonstrates that our 
methodology for characterizing subspace leakage can indeed be expected to yield 
accurate upper bounds on the subspace leakage in the vast majority of cases.

\section{Efficiency of the protocol}
\label{sec:efficiency}

The protocol presented in the previous section allows us to determine quantitative 
bounds on the subspace leakage for imperfect qubits by determining only the main 
peaks in the Fourier spectrum.  An alternative strategy is to try to identify all 
peaks in the Fourier spectrum.  The presence of any peaks in addition to the two 
main peaks is indicative of subspace leakage and a quantitative estimate of the
leakage rate can be obtained by determining the heights of the additional peaks.  
Both approaches have potential advantages and disadvantages.  The former approach 
requires only the identification of the two main peaks but these need to be clearly 
resolved and the peak heights determined with high precision.  The latter approach 
does not require precise estimates of peak heights but relies on the detection of 
additional peaks, which for high confinement will be much smaller than the major 
peaks, and are likely to be difficult to discriminate from the noise floor.  This 
raises the question which strategy is more efficient to decide if the subspace
leakage for a given qubit is below a certain error threshold.  

To answer this question, we performed a series of numerical simulations comparing 
the total number of measurements required to ascertain that the lower bound on the 
leakage rate $\epsilon_{l} = 1-\sqrt{h_0+2h_{0,1}}> 0$ within error bounds, 
versus identifying a statistically significant third peak in the Fourier spectrum,
indicating an out-of-subspace transition, for various trial Hamiltonians.  For the 
purpose of the simulations we consider the following trial Hamiltonians
\begin{equation}
H_3 = \begin{pmatrix} 0 & 1 & \gamma \\
		      1 & 1 & 0 \\
		 \gamma & 0 & 1.5
      \end{pmatrix}
\label{eq:H3}
\end{equation}
representing a system with a variable coupling $\gamma$ to a third level, as well 
as the four-level system governed by the Hamiltonian~(\ref{eq:H4}) and a six-level 
system governed by
\begin{equation}
H_6 = \begin{pmatrix} 0 & 1 & \gamma & \gamma &\gamma &\gamma \\
				1 & 1 & 0 & 0 & 0 & 0 \\
				\gamma & 0 & 1.5 & 0 & 0 & 0 \\
				\gamma & 0 & 0 & 1.7 & 0 & 0\\
				\gamma & 0 & 0 & 0 & 1.9 & 0\\
				\gamma & 0 & 0 & 0 & 0 & 2.2
\end{pmatrix},
\label{eq:H6}
\end{equation}
representing systems with variable but equal coupling to between one and four 
out-of-subspace levels, respectively.  

The lower bound, $\epsilon_l$, is taken to be non-zero for a discrete data 
set, if the analytical value $\epsilon'_l$ of the lower bound calculated 
directly from the Hamiltonian exceeds six times 
the uncertainty, $\delta(\epsilon_{l})$, for the discrete 
data calculated from the simulated Fourier spectrum, i.e.,   
\begin{equation}
\begin{aligned}
 &\epsilon'_{l} - 6\delta(\epsilon_{l}) > 0, \\
 &\delta(\epsilon_{l}) = \frac{3\delta h}{2\sqrt{h_0+2h_{0,1}}}.
 \end{aligned}
 \label{eq:in2}
\end{equation}
Six times the uncertainty in $\epsilon_{\l}$ represents the total distance
between the maximum and minimum possible value of $\epsilon_{l}$
(using a $3\sigma$ upper and lower confidence bound) and this interval should
be smaller than the analytical value, $\epsilon'_{l}$.

A peak $F(\omega')$ in the discrete Fourier spectrum is taken to be significant if 
it is more than three standard deviations $\delta h = \text{sd}[\nu({\omega})]$ 
above the projection noise floor $\bar{ \nu}(\omega)$, i.e.,
\begin{equation}
  F(\omega') - \bar{ \nu}(\omega) - 3\delta h > 0.
\label{eq:in1}
\end{equation}
This definition will underestimate the number of ensemble measurements required 
slightly as it only represents the point where the third peak is greater than at 
least 99.7\% of the noise channels.  
\begin{figure}[ht]
\includegraphics[width=0.45\textwidth]{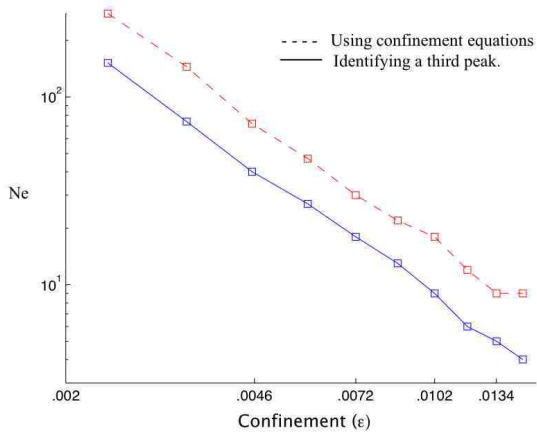}
\caption{Number of ensemble measurements required to ascertain statistically
significant subspace leakage (imperfect confinement) for the three-level system 
governed by (\ref{eq:H3}) as a function of the (analytically calculated) 
confinement using the confinement equations~(\ref{eq:in2}) and by directly 
identifying the third transition peak.}
\label{fig:co1}
\end{figure}

For the simulations a range of out-of-subspace coupling strengths $\gamma$ was 
chosen for each of the trial Hamiltonians (\ref{eq:H3}), (\ref{eq:H4}) and 
(\ref{eq:H6}), and the corresponding subspace leakage rate $\epsilon$ as well 
as the analytical lower bound $\epsilon_l'$ computed.  For each of the
Hamiltonians we then simulated experimental Rabi data and computed the discrete
Fourier spectrum.  The observation time in all cases was 30 Rabi cycles and 
the number of ensemble measurements was 
$N_e = 1024$.  The number of ensemble measurements for the
Rabi data simulations was gradually increased until a statistically significant
third peak was found~(\ref{eq:in1}), or (\ref{eq:in2}) was satisfied, respectively.

\begin{widetext}

\begin{figure}[ht]
\includegraphics[width=\textwidth]{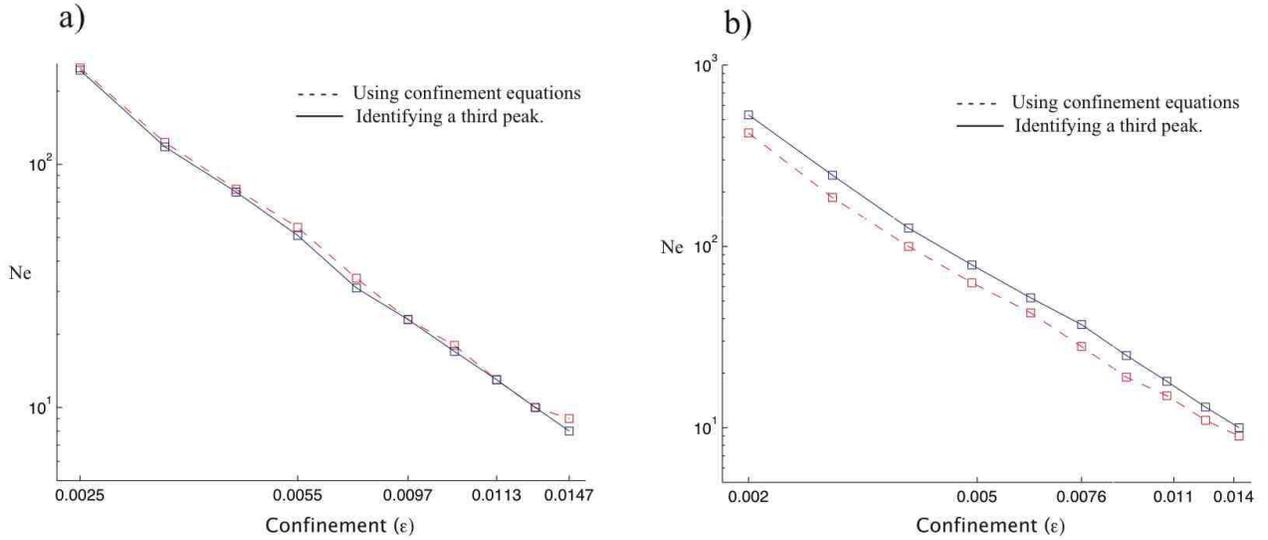}
\caption{Number of ensemble measurements required to ascertain significant subspace 
leakage (imperfect confinement) for the four-level system governed by (\ref{eq:H4}) [Fig. a] 
and the six-level system governed by (\ref{eq:H6}) [Fig. b] using the confinement equations 
and identifying a third peak.}
\label{fig:co2}
\end{figure}
\end{widetext}

Fig.~\ref{fig:co1} shows the number of ensemble measurements $N_e$ necessary
to conclude that the system is imperfect in the sense that leakage is
statistically significant for the three-level system governed by~(\ref{eq:H3})
for both methods.  The horizontal axis represents the analytical value of
confinement $\epsilon(\gamma)$.  Both curves scale roughly
$1/\sqrt{N_e}$, which is consistent with the scaling of the projection noise,
and hence the errors associated with estimating $\epsilon_l$ and
detecting a statistically significant third peak.  For the three-level system
it is clear that confirming imperfect confinement by verifying~(\ref{eq:in2})
requires more ensemble measurements than detecting a third peak according
to~(\ref{eq:in1}).  This is not too surprising since for a three-level system
there is only one additional transition $\ket{0}\leftrightarrow \ket{2}$, and
from the derivations of the confinement equations~(\ref{eq:identity}) we have,
\begin{equation}
\begin{aligned}
1 &= \bigg{(}\sum_{a=0}^{N-1}|c_a|^2\bigg{)}^2 \\
  &=  \sum_{a=0}^{N-1} |c_a|^4 + \sum_{a,b}|c_a|^2|c_b|^2 
  &= h_0 + \sum_{a,b}h_{a,b},
\end{aligned}
\end{equation}
i.e., there is a conservation law for the cumulative sum of all the peak
heights.  Hence, if the number of possible additional peaks is small, then for
a given level of confinement, the additional peaks will be greater, and thus
easier to detect, than for a system with weak coupling to a large number of
out-of-subspace levels, and hence many small transition peaks.  We therefore
conjecture that estimating subspace leakage using~(\ref{eq:in2}) will become
preferable for a system with coupling to multiple out-of-subspace levels.  The
results of numerical simulations for the Hamiltonians (\ref{eq:H4}) and (\ref{eq:H6}), shown in
Fig~\ref{fig:co2} support this conjecture.  We
observe the same general scaling behavior as for the three-level system.  For
the four-level system it is clear that although searching for the additional
transition peak is still somewhat more efficient, the difference between both
methods is small.  For the six-level the curves have swapped position, i.e.,
using the confinement equations has become a more efficient way to ascertain
statistically significant subspace leakage.

In Appendix A we have included simulations for similar Hamiltonians up to ten 
levels to show the effective crossover of the curves and how the efficiency 
difference between the two methods increases with the number of additional 
levels.  Note that for all the simulations we have endeavored to look at 
approximately the same range of subspace leakage.  From these simulations it 
is clear that searching for the third peak in the Fourier spectrum is only 
really beneficial for systems with at most one extra transition.  Hence, the
proposed method for estimating subspace leakage will be more efficient than
obvious alternatives in most cases.

\section{The effect of Decoherence}
\label{sec:decohere}

It is well known that even if subspace leakage is theoretically suppressed 
for an arbitrary control field, it is unlikely that decoherence will also be 
suppressed.  Hence, we need to examine if the proposed confinement protocol 
will still be effective in the open system case when a qubit is subject to decoherence, possibly of 
the same order, or greater, than subspace leakage.

The study of arbitrary decoherence for $N$-level systems is a lengthy discussion, 
including Markovian and possible non-Markovian processes.  Even for the simpler 
case of Markovian decoherence we would need to consider the complete $N$-level 
decoherence model with all the associated restrictions of completely positive 
maps~\cite{sonia}.  Hence, we will instead only focus on a restricted case to 
show that, for a simple example, decoherence does not invalidate the protocol.  
It should be stressed that this only represents a preliminary analysis under a 
specific model of decoherence.  Further work will involve investigating more 
complicated and system-specific decoherence effects such as $N$-level dephasing 
and spontaneous emission as well as possible system specific non-Markovian 
decoherence.  However, due to the extremely complicated nature of such an 
analysis we will limit our discussion to a specific case.

We consider a perfectly confined qubit which undergoes Markovian decoherence 
and hence can be described by the quantum Liouville equation
\begin{equation}
\partial_t \rho = -\frac{i}{\hbar}[H,\rho] + \sum_{k=1}^3 \Gamma_k \mathcal{L}_k[\rho]
\label{eq:master}    
\end{equation}
where, $\mathcal{L}_k[\rho]=([L_k,\rho L_k^{\dagger}]+[L_k\rho, L_k^{\dagger}])/2$, 
$H$ represents the single qubit control Hamiltonian, and $L_k$ are the Lindblad 
quantum jump operators, which describe the effect of the environment on the 
system, each parameterized by some rate $\Gamma_k \geq 0$.

For a basic decoherence analysis we restrict the Lindblad operators to the 
Pauli set, $\{L_k\} = \{X,Y,Z\}$, and consider a perfectly confined, control 
Hamiltonian of the form
\begin{equation}
 H = \frac{d}{2}[\cos(\theta)Z+\sin(\theta)X].
\end{equation}
This decoherence model is sufficient to describe pure dephasing as well as 
symmetric population relaxation processes in any basis, although not asymmetric 
relaxation processes.  Including each Pauli Lindblad term with an associated 
decoherence rate eliminates the problem of a preferential basis for qubit 
decoherence since any basis change of the overall system will only act to 
change the form of the Hamiltonian.

We can solve the master equation under this model by using the Bloch vector
formalism.  Expressing the density matrix as $\rho(t) = I/2+x(t)X+y(t)Y+z(t)Z$, 
Eq.~(\ref{eq:master}) takes the form $\partial_t S(t) = A S(t)$, where $S(t) 
= (x(t),y(t),z(t))^T$ and
\begin{equation}
A = \begin{pmatrix} -2(\Gamma_y+\Gamma_z) & -d \cos(\theta) & 0 \\
d \cos(\theta) & -2(\Gamma_x+\Gamma_z) & -d \sin(\theta) \\
0 & d \sin(\theta) & -2(\Gamma_x+\Gamma_y) \end{pmatrix}.
\label{eq:master1}
\end{equation}
The Rabi oscillations under this evolution are described by the function $f(t) = 
\text{Tr}[P_0 \rho(t)] = (1/2)+z(t)$, where $P_0 = |0\rangle\langle0|$, with 
an initial state $\rho(0) = |0\rangle\langle0| \implies S(0) = (0,0,1/2)^T$.  
Taking the Fourier transform of $f(t)$ leads to the rather complicated general
expression~(\ref{eq:zw}) in Appendix B.  The real component of this function 
describes three Lorentzians centered about $\omega = 0$ and $\omega=\pm d$.  
Assuming that $d \gg \Gamma_{x,y,z}$, we can expand Eq.~(\ref{eq:zw}) around 
$\omega = 0$ and $\omega = \pm d$ to obtain the functions [See Appendix B],
\begin{equation}
\begin{aligned}
h_0    &= \frac{1}{2}\delta(\omega) + 
\frac{\cos^2(\theta)}{2}\frac{\Gamma_{\alpha}}{w^2 + \Gamma_{\alpha}^{2}}, \\
h_{0,1} &= \frac{\sin^2(\theta)}{4}\frac{\Gamma_{\beta}}{(\omega \pm d)^2+\Gamma_{\beta}^2},
\end{aligned}
\label{eq:master2}
\end{equation}
where $\Gamma_{\alpha} = 2(\Gamma_y+\Gamma_z+\cos^2(\theta)(\Gamma_x-\Gamma_z))$, 
$\Gamma_{\beta} = \Gamma_x(1+\sin^2(\theta))+\Gamma_y+\Gamma_z(2-\sin^2(\theta))$ and 
$h_0$ contains a $\delta(\omega)$ offset due to the fact we are measuring the 
observable $P_0$.  
In order to describe how the maximum peak of each Lorentzian varies with $\Gamma$ 
we integrate $h_0$ and $h_{0,1}$ around an interval $\eta$ of the peak height
\begin{equation}
\begin{aligned}
h_0(\eta)    &=  \frac{\cos^2(\theta)}{2} \int_{-\eta}^\eta d\omega
\frac{\Gamma_{\alpha}}{\omega^2+\Gamma_{\alpha}^2} + \frac{1}{2}\int_{-\eta}^\eta 
d\omega \delta(\omega)\\
&= \frac{1}{2}+\frac{\cos^2(\theta)}{\pi}\arctan\bigg{(}\frac{\eta}{\Gamma_{\alpha}}\bigg{)}, \\
h_{0,1}(\eta) &= \frac{\sin^2(\theta)}{4} \int_{ d-\eta}^{d+\eta} 
d\omega \frac{\Gamma_{\beta}}{(\omega-d)^2+\Gamma_{\beta}^2} \\
&=\frac{\sin^2(\theta)}{2\pi}\arctan\bigg{(}\frac{\eta}{\Gamma_{\beta}}\bigg{)}.
\end{aligned}
\end{equation}
Hence, under decoherence the peak heights in the Fourier spectrum vary as a 
function of the integration window $\eta$ and the decoherence rates 
$\Gamma_{\alpha,\beta}$.  This is consistent since as $\Gamma_{\alpha,\beta} 
\rightarrow 0$, both $\arctan$ functions approach $\pi/2$ and $h_0+2h_{0,1} = 1$.  
The integration window $\eta$ is analogous to frequency resolution of the Fourier 
transform $\Delta \omega$, while the total area of the Lorentzian is equal to the 
peak heights when $\Gamma_{x,y,z} = 0$.  Hence for small $\Gamma_{x,y,z}$ we can 
simply choose the resolution of the Fourier transform such that the entire 
Lorentzian is essentially contained within the data channel of the primary peak.

Consider the case where we wish to ensure that the subspace leakage does not 
exceed $\zeta$.  Using the upper bound for the subspace leakage (\ref{eq:upper}) 
we have, assuming that the integration interval is approximately equal to the 
frequency resolution of the DFT (i.e. $\eta \approx \Delta \omega$)
\begin{equation}
\begin{aligned}
\zeta &= \frac{1}{2}\bigg{(}1-\sqrt{2h_0(\Delta \omega)+4h_{0,1}(\Delta \omega)-1}\bigg{)}, \\
 \frac{(1-2\zeta)^2+1}{2} &= \frac{1}{2}+\frac{\cos^2(\theta)}{\pi}\arctan\bigg{(}\frac{\Delta \omega}{\Gamma_{\alpha}}\bigg{)} \\
&\quad \quad+\frac{\sin^2(\theta)}{\pi}\arctan\bigg{(}\frac{\Delta \omega}{\Gamma_{\beta}}\bigg{)} \\
\frac{\pi(1-2\zeta)^2}{2} &= \arctan\bigg{(}\frac{\Delta \omega}{\Gamma}\bigg{)} .
\end{aligned}
\label{eq:bound}
\end{equation}
Here the last line assumes that $\Gamma_{\alpha} \approx \Gamma_{\beta} = \Gamma$.
When the Rabi frequency is much greater than the inverse of the decoherence rate 
(as necessary for any qubit realistically considered for quantum information 
processing), then the entire Lorentzian broadening caused by decoherence will be
contained within one frequency channel.  Thus, Eq.~(\ref{eq:bound}) allows us to 
calculate the maximum frequency resolution of the Fourier transform for successful 
leakage estimation using our protocol.  For example, if $\Gamma \approx 10^{-4}\mathrm{s}^{-1}$ 
and we wish to confirm that the subspace leakage is at most $\epsilon_{\text{max}} 
= 10^{-8}$, then the resolution of the Fourier transform cannot exceed $\Delta f 
\approx 250$Hz if only the primary peak channels are used.  Obviously, this 
restriction on the frequency resolution can be lifted by including multiple 
channels around the central peak when estimating the peak area.  

Although the decoherence model considered is not the most general possible case
for an imperfectly confined control Hamiltonian, this calculation demonstrates 
that the effect of decoherence does not void the protocol for estimating subspace
leakage for a common decoherence model.  A more detailed analysis considering a 
full $N$-level decoherence model, including the effect of spontaneous emission 
and absorption processes and the possibility of system-specific non-Markovian 
decoherence is desirable but beyond the scope of the current paper.

\section{Conclusions}

We have introduced an intrinsic protocol for ``quantifying" the degree of
subspace leakage for a realistic `qubit' system.  The protocol relies
on very minimal theoretical assumptions regarding qubit structure and control,
and utilizes a measurement model that is restrictive but extremely common to a
wide range of qubit systems.  We have introduced a quantitative measure of
subspace leakage, and shown that the discretization noise as a result of
finite sampling does not limit the ability of the protocol to quantify (with
appropriate error/confidence bounds) the subspace leakage for well-confined
(near perfect) qubits.

The ability to experimentally characterize subspace leakage to a high degree 
of accuracy using automated, system independent methods, which rely on the intrinsic
control and measurement apparatus of the quantum device (required for 
standard quantum information processing) 
will be vital for the commercial success of quantum nano-technology.  This 
protocol represents one of the first steps in a general library of 
characterization techniques that will be required as ``quality control'' 
protocols once mass manufacturing of qubit systems becomes common.  

Although, in this discussion, the qubit state $\ket{1}$ is only defined through the strongest 
transition it should be emphasized that if confinement estimates are made on multiple 
control fields (for example two separate Hamiltonians which induce orthogonal axis rotations), 
the computational $\ket{1}$ state {\em must} be common for both Hamiltonians.  This is 
not a significant problem, since for well engineered qubits, the computational $\ket{1}$ state 
will be known on theoretical grounds.

There are many open problems including subspace leakage estimates for systems 
undergoing a whole range of potential decoherence processes, quantifying 
confinement for multi-qubit control Hamiltonians and combining these schemes 
with other proposed methods for system characterization.  Hopefully, in 
the near future, a complete set of characterization protocols will be developed 
which will augment large scale manufacturing techniques, allowing for efficient 
and speedy transition of quantum technology from the physics laboratory to the 
commercial sector.


\appendix

\section{Efficiency comparison for leakage detection protocols}

The following simulations examined the minimal number of ensemble measurements 
required to detect imperfect qubits either via the confinement equations or by 
directly detecting the third transition peak.  Three-level, four-level and 
six-level Hamiltonians are found in the main text, the additional simulations 
were performed for all other multi-level systems up to ten levels.  The general 
form of each of the trial Hamiltonians are subsets of the ten-level system,
\begin{equation}
H_{10} = \sum_{k=0}^{9} E_k |k\rangle \langle k| + \gamma_k(|0\rangle \langle k| 
  +|k\rangle \langle0|)
\end{equation}
where $\{E_k\} \equiv \{0, 1,1.5,1.7,1.9,2.2,2.5,2.7,3,3.2\}$, $\gamma_1 = 1$ and 
$\gamma_k = \gamma$ for $k \neq 1$.

\begin{figure}[ht]
\includegraphics[width=0.45\textwidth]{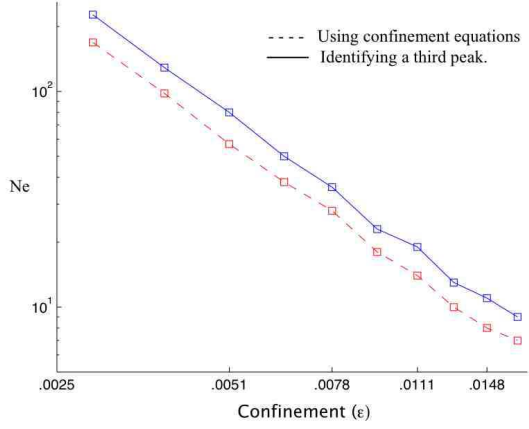}
\caption{Number of ensemble measurements required to ascertain significant subspace 
leakage (imperfect confinement) for the ten-level system using the confinement equations 
and identifying a third peak.}
\label{fig:10level}
\end{figure}

For each lower level system the appropriate Hamiltonian is simply formed by 
removing the appropriate number of rows and columns from $H_{10}$ (i.e. compare 
$H_4$ and $H_6$ in Eqs.~(\ref{eq:H4}) and (\ref{eq:H6})).  Each of these systems 
were simulated leading to the following results [Figs~\ref{fig:5level}, \ref{fig:7level} and \ref{fig:10level}], 

\begin{widetext}

\begin{figure}[ht]
\includegraphics[width=0.9\textwidth]{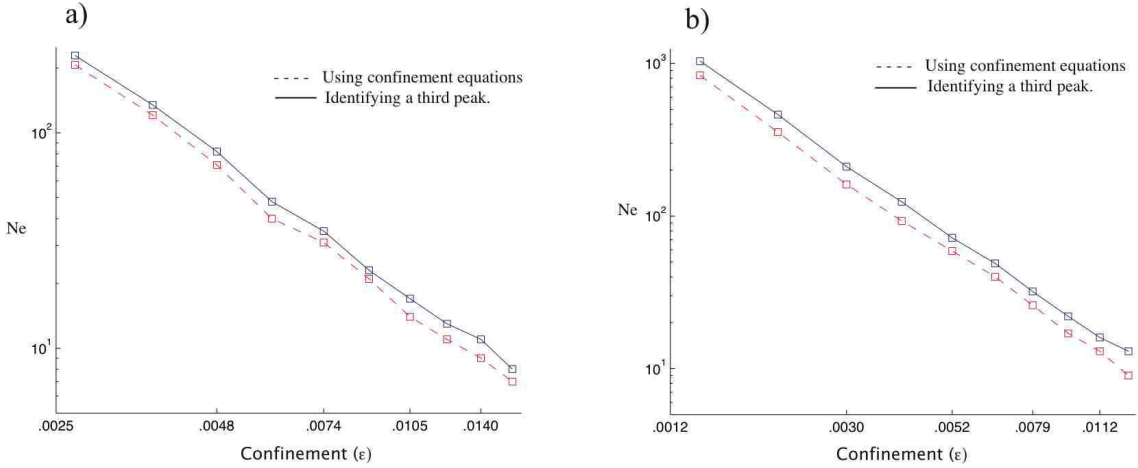}
\caption{Number of ensemble measurements required to ascertain significant subspace 
leakage (imperfect confinement) for the five-level system [Fig. a] 
and the eight-level system [Fig. b] using the confinement equations 
and identifying a third peak.}
\label{fig:5level}
\end{figure}

\begin{figure}[ht]
\includegraphics[width=0.9\textwidth]{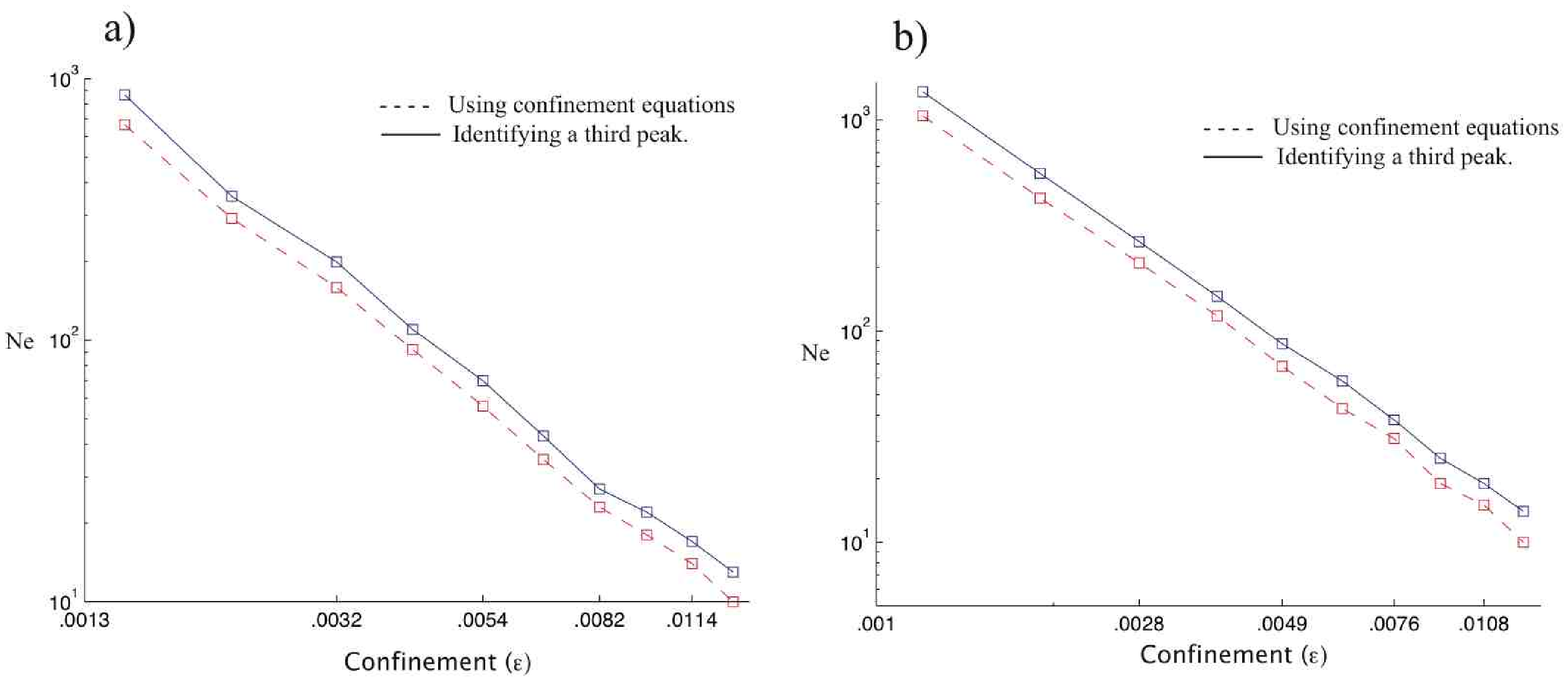}
\caption{Number of ensemble measurements required to ascertain significant subspace 
leakage (imperfect confinement) for the seven-level system [Fig. a] 
and the nine-level system [Fig. b] using the confinement equations 
and identifying a third peak.}
\label{fig:7level}
\end{figure}
\end{widetext}

\section{Solutions to the decoherence master equation}
Here we show the derivations of Eq. \ref{eq:master2} by solving the Bloch equation
$\partial_t S(t) = A S(t)$, with $A$ given in Eq.~(\ref{eq:master1}).  To solve this 
differential equation, we convert to Fourier space.  Since the Fourier transform for
a system governed by decoherence-induced semi-group dynamics is only defined for 
$t \geq 0$, we use the cosine and sine transforms
\begin{equation}
\begin{aligned}
&\mathcal{C}[f(t);\omega] = \int_0^{\infty} f(t)\cos(\omega t), \\
&\mathcal{S}[f(t);\omega] = \int_0^{\infty} f(t)\sin(\omega t), \\
\end{aligned}
\end{equation}
noting that
\begin{equation}
\mathcal{C}[f(t);\omega] - i\mathcal{S}[f(t);\omega] = \int_0^{\infty} f(t)e^{-i\omega t} = \mathcal{F}_+[f(t);\omega].
\end{equation}
Taking the sine and cosine transforms of $\partial_t S(t) = A S(t)$, noting that
\begin{equation}
\begin{aligned}
\mathcal{C}[\dot{f}(t);\omega] = \omega\mathcal{S}[f(t);\omega]-f(0), \\
\mathcal{S}[\dot{f}(t);\omega] = -\omega\mathcal{S}[f(t);\omega],
\end{aligned}
\end{equation}
gives
\begin{equation}
\begin{aligned}
\omega \mathcal{S}[S(t);\omega] - S(0) = A\mathcal{C}[S(t);w], \\
-\omega \mathcal{C}[S(t);\omega]  = A\mathcal{S}[S(t);w].
 \end{aligned}
 \end{equation}
Combining these equations and setting $S(\omega) = \mathcal{F}_+[S(t); \omega]$ 
yields,
\begin{equation}
i\omega S(\omega) - S(0) = AS(\omega),
\end{equation}
and hence
\begin{equation}
S(\omega) = -(A-i\omega I)^{-1}S(0).
\end{equation}
The initial condition $S(0)=(0, 0, 1/2)^T$ thus gives,
\begin{widetext}
\begin{equation}
FT[z(t)] = -\frac{c^2d^2+(2\Gamma_x+2\Gamma_z+i\omega)(2\Gamma_y+2\Gamma_z+i\omega)}
{2(c^2d^2+(2\Gamma_x+2\Gamma_z+i\omega)(2\Gamma_y+2\Gamma_z+i\omega))(-2(\Gamma_x+\Gamma_y)-i\omega)-2d^2s^2(2\Gamma_y+2\Gamma_z+i\omega)},
\label{eq:zw}
\end{equation}
\end{widetext}
where $c = \cos(\theta)$ and $s=\sin(\theta)$.  The subsequent expansions are too 
lengthy to include here, however standard symbolic toolkits such as Mathematica 
can handle such expressions.  The first step is to consider only the real component 
of $FT[z(t)]$.  Next, the denominator is expanded to second order around $\omega=0$ 
or $\omega = \pm d$.  After this, we expand the numerator and denominator, neglecting  
all terms of the form $\Gamma_{x,y,z}/d$ and smaller, assuming $\Gamma_{x,y,z} \ll d$ 
and being careful to note that for expansions around $\omega = \pm d$ we must keep 
terms of the form $\omega \Gamma_{x,y,z}/d$.  After simplifying the expressions we 
find
\begin{equation}
\begin{aligned}
h_0 &= \frac{\cos^2(\theta)}{2}\frac{\Gamma_{\alpha}}{w^2 + \Gamma_{\alpha}^{2}}, \\
h_{0,1} &= \frac{\sin^2(\theta)}{4}\frac{\Gamma_{\beta}}{(\omega \pm d)^2 + \Gamma_{\beta}^2},
\end{aligned}
\end{equation}
where $\Gamma_{\alpha} = 2(\Gamma_y+\Gamma_z+\cos^2(\theta)(\Gamma_x-\Gamma_z))$ and 
$\Gamma_{\beta} = \Gamma_x(1+\sin^2(\theta))+\Gamma_y+\Gamma_z(2-\sin^2(\theta))$.  
Confirming that Eq.~(\ref{eq:master2}) describes three Lorentzian curves centered 
on $\omega = 0$ and $\omega = \pm d$.

\begin{acknowledgments}
  SJD acknowledges the support of the Rae \& Edith Bennett Travelling Scholarship.
  SGS acknowledges support from an EPSRC Advanced Research Fellowship and the
  Cambridge-MIT Institute.  DKLO acknowledges support from Sidney Sussex College, 
  Cambridge and SUPA.
  SGS and DKLO also acknowledge support from the EPSRC QIP IRC (UK).
  SJD, JHC and LCLH are supported in
  part by the Australian Research Council, the Australian Government and the US
  National Security Agency (NSA), Advanced Research and Development Activity
  (ARDA), and the Army Research Office (ARO) under contract number
  W911NF-04-1-0290.
\end{acknowledgments}

\end{document}